\documentclass [12pt]{article}
\usepackage{latexsym}
\usepackage{amsmath}
\usepackage{amsfonts}
\usepackage{amssymb}
\usepackage{graphicx}
\usepackage[english]{babel}

\begin{document}

\title{On the Conductivity of a Magnetoactive Turbulent Plasma.}
\author{Otto G.Chkhetiani\thanks{%
ochkheti@mx.iki.rssi.ru}  \\
\textit{Space Research Institute, Russian Academy of Sciences,}\\
\textit{Profsoyuznaya ul. 84/32, Moscow 117997, Russia} }
\date{}\maketitle
\begin{abstract}
The problem of determining the effective conductivity tensor of a
magnetoactive turbulent plasma is considered in the approximation of
isolated particles. Additional gyrotropicterms are shown to appear in the
conductivity tensor in the presence of mean, nonzero magnetic helicity. The
dispersion of propagating electro- magnetic waves changes, additional modes
and additional rotation of the polarization plane appear, and the waves can
be amplified. The properties acquired by plasma with helicity are similar
those observed in chiral and bianisotropic electrodynamic media.
\end{abstract}

\selectlanguage{english}

\section{INTRODUCTION}

\bigskip An effective description of the propagation of waves and particles
in fluctuational magnetic fields in a turbulent conductive medium is of
great importance in solving the various problems of plasma physics and
astrophysics. The phenomena associated with the presence of small-scale
magnetic helicity $\left\langle \mathbf{AB}\right\rangle $ ($\mathbf{B}=$rot$%
\mathbf{A}$), which manifest themselves virtually on all scales of plasma
systems, play a special role here. Whereas large-scale helicity contributes
to the stability of electromagnetic structures [1], its presence at the
level of fluctuations is a nonequilibrium phenomenon that is accompanied by
various large-scale instabilities [2]. Other effects produced by small-scale
helicity, such as an asymmetry in the particle distribution and
acceleration, are well known in the diffusion theory of cosmicray
propagation [3--6]. The gyrotropic acceleration effects are also known in a
laboratory plasma as helicity input effects [7, 8]. The appearance of
additional helicity-related transport was also shown to be possible in [9].
Changes in the transport properties are also directly reflected in the
dielectric (conductive) properties of a plasma medium. Thus, for example, it
was shown in [10] that in the presence of fluctuational magnetic helicity in
the low-conductivity limit, the effective current in an isotropic plasma
proves to be dependent on the curl of the electric field ( $\mathbf{j}%
=\sigma \mathbf{E}+\sigma _{\kappa }$rot$\mathbf{E}$), which causes the mean
magnetic field to grow under certain conditions. In [10], the external
magnetic field was disregarded. In natural and laboratory conditions, the
plasma is always under the influence of large-scale magnetic fields that
affect significantly its properties [11]. Since magnetic helicity also
emerges in plasma systems in the presence of a large-scale magnetic field, a
study of its influence should take into account this factor. In [12], it was
shown for an exactly solvable model of nonlinear dynamo that the diffusion
and generation rate are strongly suppressed even in a relatively weak
magnetic field, and the regime of fast dynamo transforms into the regime of
slow dynamo with a linear growth with time. The goal of this work is to
study the effective conductivity of a turbulent magnetoactive plasma with
nonzero magnetic helicity. The kinetic approach is commonly used for a
thorough theoretical description of plasma problems. However, an allowance
for the fluctuational effects of gyrotropy is rather difficult to make and
is possible in finished form only with an appreciable number of assumptions
and simplifications (see, e.g., [6, 9]). At the same time, many basic plasma
properties can be determined in the approximation of isolated particles [11,
13], which will be used below. The statistical parameters of the
electromagnetic fluctuations are assumed to be stationary and uniform. In
Section 2, we consider the equations of motion for particles and calculate
the effective Lorentz force by the functional method with an allowance made
for the nonuniformity of the electromagnetic perturbations to within the
first order of the perturbation theory. In Section 3, we determine the
effective conductivity tensor. Fluctuational magnetic helicity gives rise to
new gyrotropic terms. Our analysis of the dispersion relation both in the
approximation of $\delta $--correlated (in time) fluctuations (Section 4)
and in the opposite case of long correlation times and high frequencies
(Section 6) for electromagnetic waves and the evolution of the magnetic
field (in the low-frequency limit) (Section 5) reveals changes in the
dispersion of propagating waves and the presence of instabilities. The
characteristic scales and growth rates of the instabilities are determined
by the relationship between the fluctuational helicity and the energy and
the external magnetic field. A magnetoactive turbulent plasma with helicity
acquires properties similar to those of chiral and bianisotropic
electrodynamic media, which have been extensively studied in recent years
[14, 15]. In Conclusions, we discuss our results and implications of the
detected effect.\bigskip

\section{BASIC EQUATIONS}

\bigskip Let us consider the motion of a one-component, singly charged
plasma in a fluctuational electromagnetic field with given correlation
properties. We will consider a cold plasma where the approximation of
isolated particles [11, 13] can be used. A regular large-scale nonuniform
perturbation of the electromagnetic field that is too weak to change
significantly the correlation properties of the electromagnetic
fluctuations, which are supposed to be given, stationary, and uniform, is
assumed to arise in the system. The expression for the electron velocity $%
\mathbf{v}$ can be written as
\begin{equation}
\frac{d\mathbf{v}}{dt}=\frac{e}{m}\left( \mathbf{E}+\frac{1}{c}\left[
\mathbf{v\times B}\right] \right) .
\end{equation}%
where $e$ and $m$ are the electron charge and mass, respectively. The
electromagnetic field and the velocity can be represented as a sum of the
large-scale slow component and the small-scale (with a zero mean) fast
component:
\begin{equation*}
\mathbf{E=}\left\langle \mathbf{E}\right\rangle +\widetilde{\mathbf{E}}%
,\quad \mathbf{B=B}_{0}+\left\langle \mathbf{B}\right\rangle +\widetilde{%
\mathbf{B}},\,\,\mathbf{v=}\left\langle \mathbf{v}\right\rangle +\widetilde{%
\mathbf{v}}\mathbf{.}
\end{equation*}%
As was said above, the mean electric and magnetic fields are assumed to be
weak compared to the fluctuational fields, i.e., $\left\langle \mathbf{E}%
\right\rangle \ll \left\langle \widetilde{\mathbf{E}}^{2}\right\rangle
^{1/2} $, $\left\langle \mathbf{B}\right\rangle \ll \left\langle \widetilde{%
\mathbf{B}}^{2}\right\rangle ^{1/2}<\mathbf{B}_{0}$. Passing to the Fourier
representation, $\left( \mathbf{F}\left( \mathbf{x},t\right) =\int \int
\widehat{\mathbf{F}}\left( \mathbf{k},w\right) \exp \left[ i\left( \mathbf{kx%
}-wt\right) \right] d\mathbf{k}dw\right) $, we write
\begin{equation}
-iw\widehat{\mathbf{v}}(\mathbf{k},w)-\frac{e}{mc}\left[ \widehat{\mathbf{v}}%
(\mathbf{k},w\mathbf{)\times B}_{0}\right] =\frac{e}{m}\widehat{\mathbf{E}}(%
\mathbf{k},w)+\frac{e}{mc}\int \left[ \widehat{\mathbf{v}}(\mathbf{q},s%
\mathbf{)\times }\widehat{\mathbf{B}}(\mathbf{k-q},w-s)\right] d\mathbf{q}ds.
\end{equation}
The equation of motion averaged over the electromagnetic fluctuations takes
the form
\begin{eqnarray}
-iw\left\langle \widehat{\mathbf{v}}(\mathbf{k},w)\right\rangle -\frac{e}{mc}%
\left[ \left\langle \widehat{\mathbf{v}}(\mathbf{k},w\mathbf{)}\right\rangle
\mathbf{\times B}_{0}\right] &=&\frac{e}{m}\left\langle \widehat{\mathbf{E}}(%
\mathbf{k},w)\right\rangle +  \notag \\
+\frac{e}{mc}\int \left[ \left\langle \widehat{\mathbf{v}}(\mathbf{q},s%
\mathbf{)}\right\rangle \mathbf{\times }\left\langle \widehat{\mathbf{B}}(%
\mathbf{k-q},w-s)\right\rangle \right] d\mathbf{q}ds &+&\frac{e}{mc}\int %
\left[ \left\langle \widehat{\widetilde{\mathbf{v}}}(\mathbf{q},s\mathbf{%
)\times }\widehat{\widetilde{\mathbf{B}}}(\mathbf{k-q},w-s)\right\rangle %
\right] d\mathbf{q}ds.  \notag \\
&&  \label{Eq0}
\end{eqnarray}
In view of the linear formulation of the problem, below we disregard the
term $\frac{e}{mc}\int \left[ \left\langle \widehat{\mathbf{v}}(\mathbf{q},s%
\mathbf{)}\right\rangle \mathbf{\times }\left\langle \widehat{\mathbf{B}}(%
\mathbf{k-q},w-s)\right\rangle \right] d\mathbf{q}ds$. The correlation $%
\left\langle \widehat{\widetilde{\mathbf{v}}}(\mathbf{q},s\mathbf{)\times }%
\widehat{\widetilde{\mathbf{B}}}(\mathbf{k-q},w-s)\right\rangle $ can be
expressed in terms of the cumulants of the fluctuational magnetic field
using the Furutsu--Novikov [16]:
\begin{eqnarray}
&&\left\langle \widehat{\widetilde{\mathbf{v}}}(\mathbf{q},s\mathbf{)\times }%
\widehat{\widetilde{\mathbf{B}}}(\mathbf{k-q},w-s)\right\rangle _{i}=  \notag
\\
&&\varepsilon _{ijk}\int \left\langle \frac{\delta \widehat{\widetilde{v}}%
_{j}(\mathbf{q},s\mathbf{)}}{\delta \widehat{\widetilde{B}}_{m}(\mathbf{k}%
^{\prime },w^{\prime })}\right\rangle \left\langle \widehat{\widetilde{B}}%
_{m}(\mathbf{k}^{\prime },w^{\prime })\widehat{\widetilde{B}}_{k}(\mathbf{k-q%
},w-s)\right\rangle d\mathbf{k}^{\prime }dw^{\prime }+  \notag \\
&&+\varepsilon _{ijk}\int \left\langle \frac{\delta ^{2}\widehat{\widetilde{v%
}}_{j}(\mathbf{q},s\mathbf{)}}{\delta \widehat{\widetilde{B}}_{m}(\mathbf{k}%
^{\prime },w^{\prime })\delta \widehat{\widetilde{B}}_{n}(\mathbf{k}^{\prime
\prime },w^{\prime \prime })}\right\rangle \times  \notag \\
&&\times \left\langle \widehat{\widetilde{B}}_{m}(\mathbf{k}^{\prime
},w^{\prime })\widehat{\widetilde{B}}_{n}(\mathbf{k}^{\prime \prime
},w^{\prime \prime })\widehat{\widetilde{B}}_{k}(\mathbf{k-q}%
,w-s)\right\rangle d\mathbf{k}^{\prime }d\mathbf{k}^{\prime \prime
}dw^{\prime }dw^{\prime \prime }+\ldots  \label{var1}
\end{eqnarray}
where the variational derivative $\displaystyle\left\langle \frac{\delta
\widehat{\widetilde{v}}_{j}(\mathbf{q},s\mathbf{)}}{\delta \widehat{%
\widetilde{B}}_{m}(\mathbf{k}^{\prime },w^{\prime })}\right\rangle $
satisfies the equation
\begin{gather}
\hat{L}_{js}\left( s\right) \left\langle \frac{\delta \widehat{\widetilde{v}}%
_{s}(\mathbf{q},s\mathbf{)}}{\delta \widehat{\widetilde{B}}_{m}(\mathbf{k}%
^{\prime },w^{\prime })}\right\rangle =-\frac{es}{mc}\varepsilon _{jlm}\frac{%
q_{l}}{q^{2}}\delta \left( s-w^{\prime }\right) \delta \left( \mathbf{q}-%
\mathbf{k}^{\prime }\right) +\frac{e}{mc}\varepsilon _{jlm}\left\langle
\widehat{v}_{l}(\mathbf{q}-\mathbf{k}^{\prime },s-w^{\prime }\mathbf{)}%
\right\rangle  \notag \\
+\frac{e}{mc}\varepsilon _{jlr}\int \left\langle \frac{\delta \widehat{%
\widetilde{v}}_{l}(\mathbf{q}^{\prime },s^{\prime }\mathbf{)}}{\delta
\widehat{\widetilde{B}}_{m}(\mathbf{k}^{\prime },w^{\prime })}\right\rangle
\left\langle \widehat{B}_{r}(\mathbf{q-q}^{\prime },s-s^{\prime
})\right\rangle d\mathbf{q}^{\prime }ds^{\prime }  \notag \\
+\frac{e}{mc}\varepsilon _{jlr}\int \left\langle \frac{\delta ^{2}\widehat{%
\widetilde{v}}_{l}(\mathbf{q}^{\prime },s^{\prime }\mathbf{)}}{\delta
\widehat{\widetilde{B}}_{m}(\mathbf{k}^{\prime },w^{\prime })\delta \widehat{%
\widetilde{B}}_{n}(\mathbf{q}^{\prime }-\mathbf{q},s^{\prime }-s)}%
\right\rangle \hat{Q}_{nr}\left( \mathbf{q}-\mathbf{q}^{\prime },s-s^{\prime
}\right) d\mathbf{q}^{\prime }ds^{\prime }.  \label{var2}
\end{gather}
Here,
\begin{eqnarray}
\hat{L}_{js}\left( s\right) &=&-is\delta _{js}+\frac{e}{mc}\varepsilon
_{jrs}B_{0r} \\
\left\langle \widehat{\widetilde{B}}_{n}(\mathbf{q-q}^{\prime
},s-s^{^{\prime }})\widehat{\widetilde{B}}_{r}(\mathbf{k},w)\right\rangle &=&%
\hat{Q}_{nr}\left( \mathbf{q-q}^{\prime },s-s^{\prime }\right) \delta \left(
\mathbf{k+q-q}^{\prime },w+s-s^{\prime }\right)  \notag
\end{eqnarray}
The second variational derivative depends on the third derivative etc. In
general, the problem is not closed. In the case of $\delta $-correlated (in
time) fluctuations, the first term is retained in Eq. (\ref{var1}), which
corresponds to the Gaussian approximation. This is also a good approximation
for short correlation times.

To take into account long correlation times, we can use, in particular, a
consistent procedure of allowance for the memory effects similar to that
suggested in [17]. Having obtained the equation for the $n-$th variational
derivative, let us substitute the emerging term with the$(n+1)-$th
derivative with an effective relaxation term, which reflects the mixing role
of the higher-order moments. This, in turn, gives rise to an effective
collision frequency determined by the pulsation amplitude of the magnetic
field in the equation for the $(n-1)-$th variational derivative, so the
frequency $s$ in an operator of the type $\hat{L}_{js}\left( s\right) $
changes to $s\prime $ $s+iw^{\ast }$. Here, we restrict our analysis to a
simpler approach and set the last term in Eq. ((\ref{var2}), as for a $%
\delta $-correlated (in time) process, equal to zero. We can verify by
direct analysis that this is possible when the characteristic frequencies of
the electromagnetic fluctuations are much higher than the stochastic Larmor
frequency determined from the mean amplitude of the magnetic fluctuations, $%
w_{fluct}\gg e\left\langle \widetilde{\mathbf{B}}^{2}\right\rangle ^{1/2}/mc$%
. This approximation is similar to the \textquotedblleft first
post-Markovian\textquotedblright\ approximation used in the statistical
theory of wave propagation in a turbulent medium [17]. Thus, for the first
variational derivative, we write%
\begin{eqnarray}
\hat{L}_{js}\left( s\right) \left\langle \frac{\delta \widehat{\widetilde{v}}%
_{s}(\mathbf{q},s\mathbf{)}}{\delta \widehat{\widetilde{B}}_{m}(\mathbf{k}%
^{\prime },w^{\prime })}\right\rangle &=&-\frac{es}{mc}\varepsilon _{jlm}%
\frac{q_{l}}{q^{2}}\delta \left( s-w^{\prime }\right) \delta \left( \mathbf{q%
}-\mathbf{k}^{\prime }\right) +\frac{e}{mc}\varepsilon _{jlm}\left\langle
\widehat{v}_{l}(\mathbf{q}-\mathbf{k}^{\prime },s-w^{\prime }\mathbf{)}%
\right\rangle  \notag \\
&&+\frac{e}{mc}\varepsilon _{jlr}\int \left\langle \frac{\delta \widehat{%
\widetilde{v}}_{l}(\mathbf{q}^{\prime },s^{\prime }\mathbf{)}}{\delta
\widehat{\widetilde{B}}_{m}(\mathbf{k}^{\prime },w^{\prime })}\right\rangle
\left\langle \widehat{B}_{r}(\mathbf{q-q}^{\prime },s-s^{\prime
})\right\rangle d\mathbf{q}^{\prime }ds^{\prime }.
\end{eqnarray}
We take into account the nonuniformity of the mean field by successive
approximations:%
\begin{eqnarray}
\hat{L}_{js}\left( s\right) ^{\prime }\left\langle \frac{\delta \widehat{%
\widetilde{v}}_{s}(\mathbf{q},s\mathbf{)}}{\delta \widehat{\widetilde{B}}%
_{m}(\mathbf{k}^{\prime },w^{\prime })}\right\rangle ^{\left( 0\right) } &=&-%
\frac{es}{mc}\varepsilon _{jlm}\frac{q_{l}}{q^{2}}\delta \left( s-w^{\prime
}\right) \delta \left( \mathbf{q}-\mathbf{k}^{\prime }\right) +\frac{e}{mc}%
\varepsilon _{jlm}\left\langle \widehat{v}_{l}(\mathbf{q}-\mathbf{k}^{\prime
},s-w^{\prime }\mathbf{)}\right\rangle  \notag \\
&& \\
\hat{L}_{js}\left( s^{\prime }\right) \left\langle \frac{\delta \widehat{%
\widetilde{v}}_{s}(\mathbf{q},s\mathbf{)}}{\delta \widehat{\widetilde{B}}%
_{m}(\mathbf{k}^{\prime },w^{\prime })}\right\rangle ^{\left( 1\right) } &=&%
\frac{e}{mc}\varepsilon _{jlr}\int \left\langle \frac{\delta \widehat{%
\widetilde{v}}_{l}(\mathbf{q}^{\prime },s^{\prime }\mathbf{)}}{\delta
\widehat{\widetilde{B}}_{m}(\mathbf{k}^{\prime },w^{\prime })}\right\rangle
^{\left( 0\right) }\left\langle \widehat{B}_{r}(\mathbf{q-q}^{\prime
},s-s^{\prime })\right\rangle d\mathbf{q}^{\prime }ds^{\prime }
\end{eqnarray}%
\bigskip

Retaining only the linear terms, we write
\begin{eqnarray}
\left\langle \frac{\delta \widehat{\widetilde{v}}_{j}(\mathbf{q},s\mathbf{)}%
}{\delta \widehat{\widetilde{B}}_{m}(\mathbf{k}^{\prime },w^{\prime })}%
\right\rangle &=&\hat{L}_{js}^{-1}\left( s\right) \left( -\frac{es}{mc}%
\varepsilon _{spm}\frac{q_{p}}{q^{2}}\delta \left( s-w^{\prime }\right)
\delta \left( \mathbf{q}-\mathbf{k}^{\prime }\right) +\frac{e}{mc}%
\varepsilon _{spm}\left\langle \widehat{v}_{p}(\mathbf{q}-\mathbf{k}^{\prime
},s-w^{\prime })\right\rangle \right) -  \notag \\
&&-\left( \frac{e}{mc}\right) ^{2}\varepsilon _{jlr}\varepsilon _{tpm}\hat{L}%
_{js}^{-1}\left( s\right) \hat{L}_{lt}^{-1}\left( w^{\prime }\right)
w^{\prime }\frac{k_{p}^{\prime }}{k^{^{\prime }2}}\left\langle \widehat{B}%
_{r}(\mathbf{q-k}^{\prime },s-w^{\prime })\right\rangle  \label{eqnv2}
\end{eqnarray}
Here, is the operator $\hat{L}_{ij}^{-1}\left( s\right) $ that is the
inverse of $\hat{L}_{ij}\left( s\right) $:
\begin{equation}
\hat{L}_{ij}^{-1}\left( s\right) =\frac{1}{is\left( s^{2}-\Omega
_{e}^{2}\right) }\left( -\delta _{ij}s^{2}+\Omega _{ei}\Omega
_{ej}-is\varepsilon _{ijk}\Omega _{ek}\right) ,\quad \mathbf{\Omega }_{e}=%
\frac{e\mathbf{B}_{0}}{mc}
\end{equation}
In what follows, we use the relationship between the fields $\mathbf{B}$ and
$\mathbf{E}$ via Maxwell's equation written in the Fourier representation as
\begin{equation}
\widehat{\mathbf{B}}\mathbf{=}\frac{c}{w}\left[ \mathbf{k\times }\widehat{%
\mathbf{E}}\right]
\end{equation}
For uniform gyrotropic fluctuations with the anisotropy introduced by a
uniform magnetic field, the correlation tensor $\widehat{Q}_{mk}(\mathbf{q}%
,s)$ is [18--20]
\begin{eqnarray}
\widehat{Q}_{mk}(\mathbf{q},s) &=&\left( \delta _{mk}-\frac{q_{m}q_{k}}{q^{2}%
}\right) \frac{E_{M}(q,(\mathbf{lq)},s)}{{4\pi q^{2}}}+i\frac{H_{M}(q,(%
\mathbf{lq)},s)}{{8\pi q^{4}}}\varepsilon _{mkt}q_{t}+  \notag \\
&&+\left[ \left( l_{m}q_{k}+l_{k}q_{m}\right) \left( \mathbf{lq}\right)
-l_{m}l_{k}q^{2}-\frac{q_{m}q_{k}}{q^{2}}\left( \mathbf{lq}\right) ^{2}%
\right] \frac{F\left( q,\left( \mathbf{lq}\right) ,s\right) }{4\pi q^{4}}-
\notag \\
&&-i\left( \delta _{ml}\varepsilon _{kij}+\delta _{kl}\varepsilon
_{mij}\right) l_{i}l_{j}\left( l_{l}q^{2}-q_{l}\left( \mathbf{lq}\right)
\right) \frac{C\left( q,\left( \mathbf{lq}\right) ,s\right) }{4\pi q^{4}}
\end{eqnarray}
Here, $\mathbf{l}$ is a unit vector parallel to the uniform magnetic field, $%
\mathbf{l}\upuparrows \mathbf{B}_{0}$. All of the correlation functions,
except for $C(q,\left( \mathbf{lq}\right) ,s)$, are even in $\left( \mathbf{%
lq}\right) $. The symmetry properties also admit the combinations linear in
components of the vector $\mathbf{l}$ considered in [6, 9]. However, it was
shown in [18, 20] that when the anisotropy is attributable to a magnetic
field, the only possible combinations are quadratic ones\footnote{%
Indeed, an arbitrary vortex field can be represented as a sum of its
toroidal and poloidal components with the basis defined for an arbitrary
direction of $\mathbf{l}$:
\begin{equation*}
h_{i}(\mathbf{x})=l_{k}\frac{\partial ^{2}P}{\partial x_{k}\partial x_{i}}%
-l_{i}\Delta P+\varepsilon _{ikj}l_{k}\frac{\partial T}{\partial x_{j}}.
\end{equation*}%
Choosing the direction of the external stationary uniform magnetic field as
this direction, we find that the dependence on the components of this
direction appears in the tensor of the pair correlations between the
magnetic fluctuations only quadratically.}. This is also confirmed by direct
calculations of the magnetic field effect on the correlation properties of
turbulence [21]. For a weak anisotropy (and for obtaining analytical
results), we can use the representation
\begin{eqnarray}
E_{M}(q,(\mathbf{lq)},s) &=&E(q,s)-\frac{(\mathbf{lq})^{2}}{q^{2}}E_{1}(q,s),
\notag \\
H_{M}(q,(\mathbf{lq)},s) &=&H(q,s)-\frac{(\mathbf{lq})^{2}}{q^{2}}H_{1}(q,s),
\notag \\
F\left( q,\left( \mathbf{lq}\right) ,s\right) &=&F\left( q,s\right) ,\
C\left( q,\left( \mathbf{lq}\right) ,s\right) =C_{1}\left( q,s\right) \left(
\mathbf{lq}\right)
\end{eqnarray}
Assuming the decay of the correlations with time to be exponential, $\sim
\frac{\tau _{\ast }}{\tau }\exp \left( -\left\vert t-t^{\prime }\right\vert
/\tau \right) $, we write for the Fourier transform
\begin{equation}
f(q,s)=f(q)\frac{\tau _{\ast }}{\pi \left( 1+s^{2}\tau ^{2}\right) }.
\label{taucor}
\end{equation}
Here, $\tau _{\ast }$ is the time constant determined by the characteristic
frequencies and scales. Thus, for example, for interplanetary plasma
turbulence [22], $\tau _{\ast }$ is assumed to be $\tau _{\ast }\sim \frac{%
\lambda }{v_{A}}=\frac{\lambda \omega _{i}}{c\Omega _{i}}$ where $\lambda $
is the characteristic fluctuational scale of the magnetic nonuniformities.
Clearly, this estimate is also valid for ionospheric plasma. Let us expand
the tensor $\widehat{Q}_{mk}(\mathbf{k-q},w-s)=\widehat{Q}_{km}(\mathbf{q-k}%
,s-w)$ as a series in $k\ll q$,
\begin{equation}
\widehat{Q}_{km}(\mathbf{q-k},s-w)=\widehat{Q}_{km}(\mathbf{q},s)-k_{r}\frac{%
\partial \widehat{Q}_{km}(\mathbf{q},s)}{{\partial }q{_{r}}}+\frac{k_{r}k_{t}%
}{2}\frac{\partial \widehat{Q}_{km}(\mathbf{q},s)}{{\partial }q{_{r}\partial
}q_{t}}+\ldots ,
\end{equation}
and substitute this representation in (\ref{eqnv2}), performing the
integration over the solid angles, the frequencies $s$. We then find that,
to within the first degree of the expansion in terms of the correlation time
$\tau $ and neglecting the effects quadratic in wave vector ($\sim k^{2}$),
the Lorentz force averaged over the uniform electromagnetic background
fluctuations is
\begin{gather*}
\frac{e}{mc}\int \left\langle \widehat{\widetilde{\mathbf{v}}}(\mathbf{q},s%
\mathbf{)\times }\widehat{\widetilde{\mathbf{B}}}(\mathbf{k-q}%
,w-s)\right\rangle d\mathbf{q}ds=-\left( \frac{e}{mc}\right) ^{2}\widehat{%
\mathcal{E}}\tau _{\ast }\left\langle \widehat{\mathbf{v}}(\mathbf{k}%
,w)\right\rangle \\
+\frac{e}{m}\left( \frac{e}{mc}\right) ^{2}\widehat{\mathcal{H}}\tau _{\ast
}i\left[ \mathbf{k\times }\left\langle \widehat{\mathbf{E}}(\mathbf{k}%
,w)\right\rangle \right] +\frac{2}{3}\left( \frac{e}{mc}\right)
^{2}E_{0}\tau _{\ast }\tau \left( 1-\frac{1}{5}t_{1}+\frac{4}{5}t_{2}\right) %
\left[ \mathbf{\Omega }_{e}\times \left\langle \widehat{\mathbf{v}}(\mathbf{k%
},w)\right\rangle \right] \\
-\frac{2}{3}\frac{e}{m}\left( \frac{e}{mc}\right) ^{2}\frac{H_{0}\tau \tau
_{\ast }}{w}\left( 1-\frac{3}{10}g_{1}\right) \left[ \mathbf{\Omega }%
_{e}\times \left[ \mathbf{k}\times \left\langle \widehat{\mathbf{E}}(\mathbf{%
k},w)\right\rangle \right] \right] -\frac{2}{3}\left( \frac{e}{mc}\right)
^{2}H_{0}\tau _{\ast }(1+i\tau w)\mathbf{\Omega }_{e}\delta ({\mathbf{k}}%
)\delta (w),
\end{gather*}%
where
\begin{eqnarray}
\left[ \widehat{\mathcal{H}}\right] _{ij} &=&\mathcal{H}_{\perp }\delta
_{ij}+\left( \mathcal{H}_{\parallel }-\mathcal{H}_{\perp }\right) l_{i}l_{j,}
\\
\mathcal{H}_{\perp } &=&H_{0}\frac{2i}{3w}\left( 1+iw\tau \right) \left( 1-%
\frac{3}{10}g_{1}\right) ,\ \mathcal{H}_{\parallel }=H_{0}\frac{2i}{3w}%
\left( 1+iw\tau \right) \left( 1-\frac{2}{5}g_{1}\right) ,
\end{eqnarray}%
\begin{eqnarray}
\left[ \widehat{\mathcal{E}}\right] _{ij} &=&\mathcal{E}_{\perp }\delta
_{ij}+\left( \mathcal{E}_{\parallel }-\mathcal{E}_{\perp }\right) l_{i}l_{j,}%
\mathcal{E}_{\perp }=\frac{4}{3}E_{0}\left( 1+i\tau w\right) \left( 1-\frac{3%
}{10}t_{1}+\frac{9}{20}t_{2}\right) , \\
\mathcal{E}_{\parallel } &=&\frac{4}{3}E_{0}\left( 1+i\tau w\right) \left( 1-%
\frac{2}{5}t_{1}+\frac{1}{10}t_{2}\right) .
\end{eqnarray}
Here,
\begin{eqnarray}
H_{0} &=&\int \frac{H\left( q\right) }{q^{2}}dq=\left\langle \widetilde{%
\mathbf{A}}\widetilde{\mathbf{B}}\right\rangle _{0};\ H_{1}=\int \frac{%
H_{1}\left( q\right) }{q^{2}}dq; \\
E_{0} &=&\int E(q)dq=\left\langle \widetilde{\mathbf{B}}^{2}\right\rangle
_{0};\ E_{1}=\int E_{1}\left( q\right) dq;\ E_{2}=\int F\left( q\right) dq.
\\
g_{1} &=&\frac{H_{1}}{H_{0}},\ t_{1}=\frac{E_{1}}{E_{0}},\ t_{2}=\frac{E_{2}%
}{E_{0}},q=\left\vert \mathbf{q}\right\vert .
\end{eqnarray}
The subscript 0 corresponds to the isotropic case. As we see, the effective
transport coefficients are directly related to the mean energy and helicity
of the fluctuational magnetic field. For the time being, let us restrict our
analysis to the approximation of a d-correlated process, $\tau \rightarrow 0$%
. The effects of finite correlation times will be considered below. For the
average Lorentz force, we then obtain
\begin{equation}
\begin{array}{c}
\frac{e}{mc}\int \left\langle \widehat{\widetilde{\mathbf{v}}}(\mathbf{q,}s%
\mathbf{)\times }\widehat{\widetilde{\mathbf{B}}}(\mathbf{k-q}%
,w-s)\right\rangle d\mathbf{q}ds= \\
\frac{e}{m}\left( \frac{e}{mc}\right) ^{2}\widehat{\mathcal{H}}_{\tau
\rightarrow 0}\tau _{1}i\left[ \mathbf{k\times }\left\langle \widehat{%
\mathbf{E}}(\mathbf{k},w)\right\rangle \right] -\left( \frac{e}{mc}\right)
^{2}\widehat{\mathcal{E}}_{\tau \rightarrow 0}\tau _{1}\left\langle \widehat{%
\mathbf{v}}(\mathbf{k},w)\right\rangle -\left( \frac{e}{mc}\right) ^{2}\frac{%
2}{3}H_{0}\tau _{1}\mathbf{\Omega }_{e}%
\end{array}
\label{delta0}
\end{equation}
The last term on the right-hand side of Eq. (\ref{delta0}) has the meaning
of constant acceleration along the external magnetic field. To all
appearances, the possibility of such acceleration was first pointed out in
[5] (see also [6]) and was also considered in detail in [7, 8] when the
helicity input was discussed. It was suggested, as an explanation, that the
acceleration is produced by the electric field generated by a fluctuational
dynamo effect. Attention to the relationship between the acceleration effect
and the transfer of electromagnetic field momentum to particles of the
medium was drawn in [23]. Assuming that $\langle \mathbf{E}\rangle ,\langle
\mathbf{B}\rangle =0$ are equal to zero, we find that in the nonrelativistic
collisionless limit, a charged particle reaches a velocity ${\mathbf{v}}%
_{max}\cong -\frac{1}{2}\left\langle \mathbf{AB}\right\rangle /\left\langle
\mathbf{B}^{2}\right\rangle \mathbf{\Omega }_{e}$ i.e., does not depend on
the correlation time and is determined by the Larmor frequency in external
magnetic field and by the scale specified by the relationship between
magnetic helicity and energy. In what follows, we disregard this effect.
This is possible for $\left\vert \mathbf{k}{\mathbf{v}}_{max}\right\vert
/\omega _{e}\ll 1$, where $\omega _{e}^{2}=\frac{4\pi ne^{2}}{m}$ and $n$ is
the electron density.

\section{THE CONDUCTIVITY TENSOR}

\bigskip Given the fluctuational friction specified by the term $%
\displaystyle-\left( \frac{e}{mc}\right) ^{2}\widehat{\mathcal{E}}_{0}\tau
_{\ast }\left\langle \widehat{\mathbf{v}}(\mathbf{k},w)\right\rangle $ the
inverse operator is
\begin{equation}
\hat{L}_{ij}^{-1}\left( w\right) =\frac{-\delta _{ij}\left( w+i\overline{%
\Omega _{\perp }^{2}}_{e}\tau _{\ast }\right) ^{2}+\Omega _{ei}\Omega
_{ej}-i\left( w+i\overline{\Omega _{\parallel }^{2}}_{e}\tau _{\ast }\right)
\varepsilon _{ijk}\Omega _{ek}}{i\left( w+i\overline{\Omega _{\parallel }^{2}%
}_{e}\tau _{\ast }\right) \left( \left( w+i\overline{\Omega _{\perp }^{2}}%
_{e}\tau _{\ast }\right) ^{2}-\Omega _{e}^{2}\right) }
\end{equation}
Here $\overline{\Omega _{\perp }^{2}}_{e}=\left( \frac{e}{mc}\right) ^{2}%
\mathcal{E}_{0\perp },\overline{\Omega _{\parallel }^{2}}_{e}=\left( \frac{e%
}{mc}\right) ^{2}\mathcal{E}_{0\parallel }$. Taking into account the
explicit form of the tensor $\widehat{\mathcal{H}}_{0}$ , let us write the
electron velocity as
\begin{gather}
\left\langle \widehat{\mathbf{v}}(\mathbf{k},w)\right\rangle =-\frac{e}{m}%
\frac{\left\langle \widehat{\mathbf{E}}(\mathbf{k},w)\right\rangle }{%
iw_{e\parallel }^{\prime }}+\frac{e}{m}\frac{\Omega _{e}^{2}\left( \mathbf{l}%
\left( \mathbf{l}\left\langle \widehat{\mathbf{E}}(\mathbf{k}%
,w)\right\rangle \right) -\left\langle \widehat{\mathbf{E}}(\mathbf{k}%
,w)\right\rangle \right) }{iw_{e\parallel }^{\prime }\left( w_{e\perp
}^{\prime 2}-\Omega _{e}^{2}\right) }  \notag \\
+\frac{e}{m}\frac{\Omega _{e}\left[ \mathbf{l}\times \left\langle \widehat{%
\mathbf{E}}(\mathbf{k},w)\right\rangle \right] }{\left( w_{e\perp }^{\prime
2}-\Omega _{e}^{2}\right) }-\frac{e}{m}\left( \frac{e}{mc}\right) ^{2}%
\mathcal{H}_{0\perp }\tau _{\ast }\frac{i}{w}\frac{w_{e\perp }^{\prime
2}\left( \left[ \mathbf{k\times E}\left( \mathbf{k}\right) \right] -\mathbf{l%
}\left( \mathbf{l}\left[ \mathbf{k\times }\left\langle \widehat{\mathbf{E}}(%
\mathbf{k},w)\right\rangle \right] \right) \right) }{w_{e\parallel }^{\prime
}\left( w_{e\perp }^{\prime 2}-\Omega _{e}^{2}\right) }  \notag \\
-\frac{e}{m}\left( \frac{e}{mc}\right) ^{2}\mathcal{H}_{0\parallel }\tau
_{\ast }\frac{i}{w}\frac{\mathbf{l}\left( \mathbf{l}\left[ \mathbf{k\times }%
\left\langle \widehat{\mathbf{E}}(\mathbf{k},w)\right\rangle \right] \right)
}{w_{e\parallel }^{\prime }}-\frac{e}{m}\left( \frac{e}{mc}\right) ^{2}%
\mathcal{H}_{0\perp }\tau _{\ast }\Omega _{e}\frac{\left[ \mathbf{l}\times %
\left[ \mathbf{k\times }\left\langle \widehat{\mathbf{E}}(\mathbf{k}%
,w)\right\rangle \right] \right] }{w\left( w_{e\perp }^{\prime 2}-\Omega
_{e}^{2}\right) },
\end{gather}
where $\Omega _{e}=\frac{e\left\vert \mathbf{B}\right\vert _{0}}{mc}%
,w_{e\perp \left( \parallel \right) }^{\prime }=w+i\overline{\Omega _{\perp
\left( \parallel \right) }^{2}}\tau _{\ast }$. The calculations for ions are
similar, and the ion velocity can be written as%
\begin{gather}
\left\langle \widehat{\mathbf{v}}(\mathbf{k},w)\right\rangle _{i}=\frac{e}{M}%
\frac{\left\langle \widehat{\mathbf{E}}(\mathbf{k},w)\right\rangle }{%
iw_{i\parallel }^{\prime }}-\frac{e}{M}\frac{\Omega _{i}^{2}\left( \mathbf{l}%
\left( \mathbf{l}\left\langle \widehat{\mathbf{E}}(\mathbf{k}%
,w)\right\rangle \right) -\left\langle \widehat{\mathbf{E}}(\mathbf{k}%
,w)\right\rangle \right) }{iw_{i\parallel }^{\prime }\left( w_{i\perp
}^{\prime 2}-\Omega _{i}^{2}\right) }  \notag \\
+\frac{e}{M}\frac{\Omega _{i}\left[ \mathbf{l}\times \left\langle \widehat{%
\mathbf{E}}(\mathbf{k},w)\right\rangle \right] }{\left( w_{i\perp }^{\prime
2}-\Omega _{i}^{2}\right) }+\frac{e}{M}\left( \frac{e}{Mc}\right) ^{2}%
\mathcal{H}_{0\perp }\tau _{\ast }\frac{i}{w}\frac{w_{i\perp }^{\prime
2}\left( \left[ \mathbf{k\times }\left\langle \widehat{\mathbf{E}}(\mathbf{k}%
,w)\right\rangle \right] -\mathbf{l}\left( \mathbf{l}\left[ \mathbf{k\times }%
\left\langle \widehat{\mathbf{E}}(\mathbf{k},w)\right\rangle \right] \right)
\right) }{w_{i\parallel }^{\prime }\left( w_{i\perp }^{\prime 2}-\Omega
_{i}^{2}\right) }  \notag \\
+\frac{e}{M}\left( \frac{e}{Mc}\right) ^{2}\mathcal{H}_{0\parallel }\tau
_{\ast }\frac{i}{w}\frac{\mathbf{l}\left( \mathbf{l}\left[ \mathbf{k\times }%
\left\langle \widehat{\mathbf{E}}(\mathbf{k},w)\right\rangle \right] \right)
}{w_{i\parallel }^{\prime }}-\frac{e}{M}\left( \frac{e}{Mc}\right) ^{2}%
\mathcal{H}_{0\perp }\tau _{\ast }\Omega _{i}\frac{\left[ \mathbf{l}\times %
\left[ \mathbf{k\times }\left\langle \widehat{\mathbf{E}}(\mathbf{k}%
,w)\right\rangle \right] \right] }{w\left( w_{i\perp }^{\prime 2}-\Omega
_{i}^{2}\right) }
\end{gather}

The subscript $i$ refers to the ion analogues of the parameters introduced
for electrons: $\Omega _{i}=\frac{e\left\vert \mathbf{B}\right\vert _{0}}{Mc}%
,w_{i\perp \left( \parallel \right) }^{\prime }=w+i\overline{\Omega _{\perp
\left( \parallel \right) }^{2}}_{i}\tau _{\ast }$. As we see, averaging over
the electromagnetic fluctuations is equivalent , in particular, to an
effective collision with frequencies proportional to $\sim \overline{\Omega
_{\perp }^{2}}_{e\left( i\right) }\tau _{\ast }$ and $\overline{,\Omega
_{\parallel }^{2}}_{e\left( i\right) }\tau _{\ast }$. For the conductivity
tensor $j_{k}=\widehat{\sigma }_{kl}(\mathbf{k},w)E_{l}(\mathbf{k},w)$ $%
\left( \mathbf{j}=ne\left( \left\langle \mathbf{v}\right\rangle
-\left\langle \mathbf{v}\right\rangle _{i}\right) \right) $, we obtain
\begin{gather}
4\pi \widehat{\sigma }_{kl}(\mathbf{k},w)=-\left( \frac{\omega
_{e}^{2}w_{e\perp }^{\prime 2}}{iw_{e\parallel }^{\prime }\left( w_{e\perp
}^{\prime 2}-\Omega _{e}^{2}\right) }+\frac{\omega _{i}^{2}w_{i\perp
}^{\prime 2}}{iw_{i\parallel }^{\prime }\left( w_{i\perp }^{\prime 2}-\Omega
_{i}^{2}\right) }\right) \delta _{kl}  \notag \\
+\left( \frac{\omega _{e}^{2}\Omega _{e}^{2}}{iw_{e\parallel }^{\prime
}\left( w_{e\perp }^{\prime 2}-\Omega _{e}^{2}\right) }+\frac{\omega
_{i}^{2}\Omega _{i}^{2}}{iw_{i\parallel }^{\prime }\left( w_{i\perp
}^{\prime 2}-\Omega _{i}^{2}\right) }\right) l_{k}l_{l}+\left( \frac{\omega
_{e}^{2}\Omega _{e}}{\left( w_{e\perp }^{\prime 2}-\Omega _{e}^{2}\right) }-%
\frac{\omega _{i}^{2}\Omega _{i}}{\left( w_{i\perp }^{\prime 2}-\Omega
_{i}^{2}\right) }\right) \varepsilon _{kml}l_{m}  \notag \\
+\left( \frac{\omega _{e}^{2}h_{\perp e}\Omega _{e}}{w\left( w_{e\perp
}^{\prime 2}-\Omega _{e}^{2}\right) }-\frac{\omega _{i}^{2}h_{\perp i}\Omega
_{i}}{w\left( w_{i\perp }^{\prime 2}-\Omega _{i}^{2}\right) }\right) \left(
l_{m}k_{m}\delta _{kl}-l_{l}k_{k}\right) -i\left( \frac{\omega
_{e}^{2}h_{\parallel e}}{ww_{e\parallel }^{\prime }}+\frac{\omega
_{i}^{2}h_{\parallel ei}}{ww_{i\parallel }^{\prime }}\right)
l_{k}l_{m}\varepsilon _{mnl}k_{n}  \notag \\
-i\left( \frac{\omega _{e}^{2}h_{\parallel e}w_{e\perp }^{\prime }}{w\left(
w_{e\perp }^{\prime 2}-\Omega _{e}^{2}\right) }+\frac{\omega
_{i}^{2}h_{\parallel ei}w_{i\perp }^{\prime }}{w\left( w_{i\perp }^{\prime
2}-\Omega _{i}^{2}\right) }\right) \left( \varepsilon
_{kml}k_{m}-l_{k}l_{m}\varepsilon _{lmn}k_{n}\right)  \label{sigma}
\end{gather}
where $\omega _{e\left( i\right) }^{2}=\frac{4\pi ne^{2}}{m\left( M\right) }$
.

The coefficients$h_{\perp e\left( i\right) }$ and $h_{\parallel e\left(
i\right) }$ have the dimensions of velocity, and it is convenient to
represent them as
\begin{equation}
\begin{array}{c}
h_{\perp e\left( i\right) }=\overline{\Omega _{\perp e\left( i\right) }^{2}}%
\tau _{\ast }\lambda _{\kappa \perp }=\alpha _{\perp e\left( i\right) }\frac{%
\Omega _{e\left( i\right) }}{\Omega _{\kappa \perp }}c,\alpha _{\perp
e\left( i\right) }=\frac{\overline{\Omega _{\perp e\left( i\right) }^{2}}%
\tau _{\ast }}{\Omega _{e\left( i\right) }}, \\
\Omega _{\kappa \perp }=\frac{c}{\lambda _{\kappa \perp }} \\
h_{\parallel e\left( i\right) }=\overline{\Omega _{\parallel e\left(
i\right) }^{2}}\tau _{\ast }\lambda _{\kappa \parallel }=\alpha _{\parallel
e\left( i\right) }\frac{\Omega _{e\left( i\right) }}{\Omega _{\kappa
\parallel }}c,\alpha _{\parallel e\left( i\right) }=\frac{\overline{\Omega
_{\perp e\left( i\right) }^{2}}\tau _{\ast }}{\Omega _{e\left( i\right) }},
\\
\Omega _{\kappa \parallel }=\frac{c}{\lambda _{\kappa \parallel }}%
\end{array}%
\end{equation}%
where the scale $\lambda _{\kappa \perp \left( \parallel \right) }$ is
defined by the ratio of the helicity and energy of the fluctuations:%
\footnote{%
The characteristic scale of the fluctuational magnetic helicity is known for
the solar-wind turbulence [24], where it lies within the range $0.004\div
0.02$ AU ($\sim $ $6\cdot 10^{8}\div 3\cdot 10^{9}$ m ).}
\begin{equation}
\lambda _{\kappa \perp \left( \parallel \right) }=\frac{\mathcal{H}_{0\perp
\left( \parallel \right) }}{\mathcal{E}_{0\perp \left( \parallel \right) }}%
\approx \frac{1}{2}\frac{\left\langle \mathbf{AB}\right\rangle }{%
\left\langle \mathbf{B}^{2}\right\rangle }.
\end{equation}%
Neglecting the fluctuational damping $\overline{\Omega _{\parallel }^{2}}%
\tau _{\ast }\>(\overline{\Omega _{\perp }^{2}}\tau _{\ast })$, we obtain
\begin{align}
4\pi \widehat{\sigma }_{kl}(\mathbf{k},w)& =iw\left( \frac{\omega _{e}^{2}}{%
w^{2}-\Omega _{e}^{2}}+\frac{\omega _{i}^{2}}{w^{2}-\Omega _{i}^{2}}\right)
\delta _{kl}+\left( \frac{\omega _{e}^{2}\Omega _{e}^{2}}{w^{2}-\Omega
_{e}^{2}}+\frac{\omega _{i}^{2}\Omega _{i}^{2}}{w^{2}-\Omega _{i}^{2}}%
\right) \frac{l_{k}l_{l}}{iw}  \notag \\
& +\left( \frac{\omega _{e}^{2}\Omega _{e}}{w^{2}-\Omega _{e}^{2}}-\frac{%
\omega _{i}^{2}\Omega _{i}}{w^{2}-\Omega _{i}^{2}}\right) \varepsilon
_{kml}l_{m}  \notag \\
& +\left( \frac{\omega _{e}^{2}\Omega _{e}h_{\perp e}}{w\left( w^{2}-\Omega
_{e}^{2}\right) }-\frac{\omega _{i}^{2}\Omega _{i}h_{\perp i}}{w\left(
w^{2}-\Omega _{i}^{2}\right) }\right) \left( l_{m}k_{m}\delta
_{kl}-l_{l}k_{k}\right)   \notag \\
& -i\left( \frac{\omega _{e}^{2}h_{\parallel e}}{w^{2}}+\frac{\omega
_{i}^{2}h_{\parallel i}}{w^{2}}\right) l_{k}l_{m}\varepsilon
_{mnl}k_{n}-i\left( \frac{\omega _{e}^{2}h_{\perp e}}{w^{2}-\Omega _{e}^{2}}+%
\frac{\omega _{i}^{2}h_{\perp i}}{w^{2}-\Omega _{i}^{2}}\right) \left(
\varepsilon _{kml}k_{m}-l_{k}l_{m}\varepsilon _{lmn}k_{n}\right) .
\end{align}%
Hence, the permittivity tensor is
\begin{eqnarray}
\varepsilon _{ij} &=&\delta _{ij}+\frac{4\pi i}{w}\widehat{\sigma }_{ij}
\notag \\
\widehat{\varepsilon } &=&\left(
\begin{array}{ccc}
\varepsilon _{\perp }+i\chi _{0}k_{z} & ig+\chi _{\perp }k_{z} & -i\chi
_{0}k_{x}-\chi _{\perp }k_{y} \\
-ig-\chi _{\perp }k_{z} & \varepsilon _{\perp }+i\chi _{0}k_{z} & -i\chi
_{0}k_{y}+\chi _{\perp }k_{x} \\
\chi _{\parallel }k_{y} & -\chi _{\parallel }k_{x} & \varepsilon _{\parallel
}%
\end{array}%
\right)   \label{epsilon}
\end{eqnarray}%
where
\begin{eqnarray}
\varepsilon _{\perp } &=&1-\frac{\omega _{e}^{2}}{w^{2}-\Omega _{e}^{2}}-%
\frac{\omega _{i}^{2}}{w^{2}-\Omega _{i}^{2}},\varepsilon _{\parallel }=1-%
\frac{\omega _{e}^{2}}{w^{2}}-\frac{\omega _{i}^{2}}{w^{2}},g=\frac{\omega
_{e}^{2}\Omega _{e}}{w\left( \Omega _{e}^{2}-w^{2}\right) }-\frac{\omega
_{i}^{2}\Omega _{i}}{w\left( \Omega _{i}^{2}-w^{2}\right) }, \\
\chi _{0} &=&\frac{\omega _{e}^{2}\Omega _{e}}{w^{2}\left( w^{2}-\Omega
_{e}^{2}\right) }h_{\perp e}-\frac{\omega _{i}^{2}\Omega _{i}}{w^{2}\left(
w^{2}-\Omega _{i}^{2}\right) }h_{\perp i}, \\
\chi _{\perp } &=&\frac{h_{\perp e}}{w}\frac{\omega _{e}^{2}}{w^{2}-\Omega
_{e}^{2}}+\frac{h_{\perp i}}{w}\frac{\omega _{i}^{2}}{w^{2}-\Omega _{i}^{2}}%
,\chi _{\parallel }=\frac{h_{\parallel e}}{w}\frac{\omega _{e}^{2}}{%
w^{2}-\Omega _{e}^{2}}+\frac{h_{\parallel i}}{w}\frac{\omega _{i}^{2}}{%
w^{2}-\Omega _{i}^{2}}
\end{eqnarray}%
As we see from (\ref{epsilon}), fluctuational helicity gives rise to
additional gyrotropic terms in the permittivity tensor. To elucidate their
role, let us analyze the dispersion relation for electromagnetic waves.

\section{THE DISPERSION RELATION}

\bigskip Denote the angle between the vectors $\mathbf{n}$ and $\mathbf{B}%
_{0}$ by $\theta $. The dispersion relation for the complex refractive index
$\mathbf{n}=c\mathbf{k}/w$ is defined as [11]
\begin{equation}
\det \left\Vert n^{2}\delta _{ij}-n_{i}n_{j}-\widehat{\varepsilon }%
_{ij}\right\Vert =0
\end{equation}
We also assume that
\begin{equation*}
\mathbf{n=}\left( n\sin \left( \theta \right) ,0,n\cos \left( \theta \right)
\right) .
\end{equation*}%
The dispersion relation is then
\begin{equation}
\begin{array}{c}
\left( g^{2}+\left( {n}^{2}-\varepsilon _{\perp }\right) \,\varepsilon
_{\perp }\right) \varepsilon _{\parallel }-{n}^{2}\,\varepsilon _{\parallel
}\,\,\left( {n}^{2}-\varepsilon _{\perp }\right) \,{\cos }^{2}{(\theta )}-{n}%
^{2}\,\left( g^{2}+\left( {n}^{2}-\varepsilon _{\perp }\right) \,\varepsilon
_{\perp }\right) {\sin }^{2}{(\theta )} \\
-{n}^{2}\,\varepsilon _{\parallel }\,w^{2}\,\left( -{\chi }_{0}^{2}+{\chi }%
_{\perp }{}^{2}\right) /c^{2}\,{\cos }^{2}{(\theta )}-{n}^{2}\,w^{2}\,\left(
g\,\chi _{0}+\varepsilon _{\perp }\,{\chi }_{\perp }\right) \,{\chi }%
_{\parallel }{/c}^{2}{\sin }^{2}{(\theta )}\, \\
+i\,\,{n}\,w\,\cos (\theta )\,\varepsilon _{\parallel }\,\left( {n}%
^{2}\,\chi _{0}-2\,\varepsilon _{\perp }\,\chi _{0}-2\,g\,{\chi }_{\perp
}\right) /c \\
+i\,\,{n}\,^{3}w\,\cos (\theta )\,\left( \,\,\,\varepsilon _{\parallel
}\,\chi _{0}/c\,{\cos }^{2}{(\theta )}+\,\left( \varepsilon _{\perp }\,\chi
_{0}+g\,{\chi }_{\perp }-g\,{\chi }_{\parallel }\right) /c\,{\sin }^{2}{%
(\theta )}\right) =0%
\end{array}
\label{dispeq}
\end{equation}%
Let us consider the waves that propagate along the magnetic field, $\theta
=0 $. In this case, the dispersion relation has the solutions
\begin{equation}
\begin{array}{c}
n_{1,2}=\frac{1}{2}\left( iw\left( \chi _{0}+\chi _{\perp }\right) /c\pm
\left( 4\left( \varepsilon _{\perp }-g\right) -w^{2}\left( \chi _{0}+\chi
_{\perp }\right) ^{2}/c^{2}\right) ^{1/2}\right) \\
n_{3,4}=\frac{1}{2}\left( iw\left( \chi _{0}-\chi _{\perp }\right) /c\pm
\left( 4\left( \varepsilon _{\perp }+g\right) -w^{2}\left( \chi _{0}-\chi
_{\perp }\right) ^{2}/c^{2}\right) ^{1/2}\right)%
\end{array}%
\end{equation}
Given that%
\begin{equation}
\begin{array}{c}
\varepsilon _{\perp }\mp g=1-\frac{\omega _{e}^{2}}{w\left( w\pm \Omega
_{e}\right) }-\frac{\omega _{i}^{2}}{w\left( w\mp \Omega _{i}\right) }, \\
\chi _{0}\pm \chi _{\perp }=\pm \alpha _{\perp e}\frac{\omega _{e}^{2}\Omega
_{e}}{w^{2}\left( w\mp \Omega _{e}\right) \Omega _{\kappa \perp }}c\pm
\alpha _{\perp i}\frac{\omega _{i}^{2}\Omega _{i}}{w^{2}\left( w\pm \Omega
_{i}\right) \Omega _{\kappa \perp }}c,%
\end{array}%
\end{equation}
the wave vector is
\begin{eqnarray*}
ck &=&i\left( \pm \frac{\alpha _{\perp e}}{2\Omega _{\kappa \perp }}\frac{%
\omega _{e}^{2}\Omega _{e}}{\left( w\mp \Omega _{e}\right) }\pm \frac{\alpha
_{\perp i}}{2\Omega _{\kappa \perp }}\frac{\omega _{i}^{2}\Omega _{i}}{%
\left( w\pm \Omega _{i}\right) }\right) \pm \\
&&\left( w^{2}\left( 1-\frac{\omega _{e}^{2}}{w\left( w\pm \Omega
_{e}\right) }-\frac{\omega _{i}^{2}}{w\left( w\mp \Omega _{i}\right) }%
\right) -\left( \pm \frac{\alpha _{\perp e}}{2\Omega _{\kappa \perp }}\frac{%
\omega _{e}^{2}\Omega _{e}}{\left( w\mp \Omega _{e}\right) }\pm \frac{\alpha
_{\perp i}}{2\Omega _{\kappa \perp }}\frac{\omega _{i}^{2}\Omega _{i}}{%
\left( w\pm \Omega _{i}\right) }\right) ^{2}\right) ^{1/2},
\end{eqnarray*}%
whence the equation for the frequency is
\begin{equation}
w^{2}-w\left( \frac{\omega _{e}^{2}}{\left( w\pm \Omega _{e}\right) }+\frac{%
\omega _{i}^{2}}{\left( w\mp \Omega _{i}\right) }\right) \pm i\frac{ck}{%
\Omega _{\kappa \perp }}\left( \pm \alpha _{\perp e}\frac{\omega
_{e}^{2}\Omega _{e}}{\left( w\mp \Omega _{e}\right) }\pm \alpha _{\perp i}%
\frac{\omega _{i}^{2}\Omega _{i}}{\left( w\pm \Omega _{i}\right) }\right)
=c^{2}k^{2}
\end{equation}%
\bigskip

At low frequencies with $\frac{\omega _{e}^{2}}{\Omega _{e}^{2}}\ll \frac{%
\omega _{i}^{2}}{\Omega _{i}^{2}}$and $\omega _{e}^{2}\gg \omega _{i}^{2}$,
the square of the frequency is
\begin{equation}
w^{2}=\frac{v_{A}^{2}k^{2}}{\left( 1+\left( 1+\alpha _{\perp i}\right) \frac{%
v_{A}^{2}}{c^{2}}\right) }\left( 1\pm i\alpha _{\perp e}k\lambda _{\kappa
\perp }\frac{\omega _{e}^{2}}{c^{2}k^{2}}\right) ,
\end{equation}
where $v_{A}^{2}=\frac{B_{0}^{2}}{4\pi nM}$. At low values of $\alpha
_{\perp e}k\lambda _{\kappa \perp }\frac{\omega _{e}^{2}}{c^{2}k^{2}}$ (for
small scales),%
\begin{equation}
w=\left( \frac{v_{A}k}{\left( 1+\left( 1+\alpha _{\perp i}\right)
v_{A}^{2}/c^{2}\right) ^{1/2}}\pm i\frac{v_{A}/c}{\left( 1+\left( 1+\alpha
_{\perp i}\right) v_{A}^{2}/c^{2}\right) ^{1/2}}\frac{\alpha _{\perp
e}\lambda _{\kappa \perp }\omega _{e}^{2}}{2c}\right) .
\end{equation}
In this case, the coefficient of the complex refractive index does not
depend on the wave vector. In contrast, at high values of $\alpha _{\perp
e}k\lambda _{\kappa \perp }\frac{\omega _{e}^{2}}{c^{2}k^{2}}$ (for large
scales),
\begin{equation}
w=\frac{v_{A}/c}{\left( 1+\left( 1+\alpha _{\perp i}\right)
v_{A}^{2}/c^{2}\right) ^{1/2}}\left( \frac{\alpha _{\perp e}k\lambda
_{\kappa \perp }}{2}\right) ^{1/2}\omega _{e}\left( 1\pm i\right) .
\end{equation}%
As we see, in the presence of magnetic fluctuation helicity, there is an
instability and the amplitude of the electromagnetic waves propagating in a
plasma increases. This demonstrates the nonequilibrium existence of
reflectional symmetry breaking at the level of fluctuations. Thus, for
example, helicity also leads to an instability, an inverse energy cascade,
in magnetohydrodynamics [2]. Unstable waves have nonzero helicity, i.e., a
vortex component of the electric field. The motion of charged particles in a
magnetic field with fluctuational helicity is equivalent to the motion in
random helical magnetic fields with preferred helix orientation.

The resonance condition during the motion of particles in a helical magnetic
field is satisfied for the particles that move in a direction opposite to
the field ($\mathbf{Bv}<0$) [25]. After averaging, this resonance condition
will correspond to the following: when the helicities of the perturbations
and fluctuations have opposite signs, the perturbations will give up energy
to particles of the medium; in contrast, when the helicities of the
perturbations and fluctuations have the same signs, the field will be
amplified --- take away energy from particles of the medium. Indeed, the
helicity of growing waves coincides in sign with the small-scale
fluctuational helicity. In the opposite case, the perturbation is damped.
Note also that the dispersion of the propagating waves changes as well. For
large scales, $w\sim k^{1/2}$, the dispersion law is similar to that of
gravity waves in deep water whose phase velocity increases with scale. Such
long waves can be revealed in the spectrum of geoelectromagnetic
perturbations. Note that the fast large--scale electric perturbations in the
E region of the ionosphere that accompany such catastrophic events as
magnetic storms and substorms, earthquakes, and man-made explosions are, to
all appearances, of a vortex nature [26]. Let us consider the range of
helicon frequencies: $\Omega _{i}\ll w\ll \Omega _{e},\omega _{e}^{2}\gg
w\Omega _{e}$. In this case, the frequency can be expressed as
\begin{equation}
w=\pm \Omega _{e}\frac{c^{2}k^{2}}{\omega _{e}^{2}}\mp i\alpha _{\perp
e}k\lambda _{\kappa \perp }\Omega _{e}
\end{equation}
The wave propagation is also accompanied by an instability with the growth
rate $\alpha _{\perp e}k\lambda _{\kappa \perp }\Omega _{e}$. Retaining the
quadratic terms in the expansion of the Lorentz force $\left\langle \widehat{%
\widetilde{\mathbf{v}}}(\mathbf{q},s\mathbf{)\times }\widehat{\widetilde{%
\mathbf{B}}}(\mathbf{k-q},w-s)\right\rangle $ in terms of large scales $%
\left( k\ll q\right) $ yields a lower limit for such instability [10], and
the perturbations are damped at $k>k_{crit}$. Let us consider the waves that
propagate perpendicular to the magnetic field, $\theta =\frac{\pi }{2}$. In
this case, the square of the complex refractive index is
\begin{equation}
\begin{array}{c}
n_{1}^{2}=\frac{\varepsilon _{\perp }\left( \varepsilon _{\perp
}+\varepsilon _{\parallel }\right) -g^{2}-\kappa +\left( \left( \varepsilon
_{\perp }\left( \varepsilon _{\perp }-\varepsilon _{\parallel }\right)
-g^{2}\right) ^{2}-2\left( \varepsilon _{\perp }\left( \varepsilon _{\perp
}+\varepsilon _{\parallel }\right) -g^{2}\right) \kappa +\kappa ^{2}\right)
^{1/2}}{2\varepsilon _{\perp }} \\
n_{2}^{2}=\frac{\varepsilon _{\perp }\left( \varepsilon _{\perp
}+\varepsilon _{\parallel }\right) -g^{2}-\kappa -\left( \left( \varepsilon
_{\perp }\left( \varepsilon _{\perp }-\varepsilon _{\parallel }\right)
-g^{2}\right) ^{2}-2\left( \varepsilon _{\perp }\left( \varepsilon _{\perp
}+\varepsilon _{\parallel }\right) -g^{2}\right) \kappa +\kappa ^{2}\right)
^{1/2}}{2\varepsilon _{\perp }} \\
\kappa =w^{2}\left( g\chi _{0}+\varepsilon _{\perp }\chi _{\perp }\right)
\chi _{\parallel }/c^{2}%
\end{array}
\label{Bperp}
\end{equation}
In the absence of helicity, the first and second expressions in (\ref{Bperp}%
) would correspond to the extraordinary and ordinary waves, respectively. As
we see, their propagation conditions change, and elliptical polarization
attributable to helicity appears in both types of waves.

\section{THE OHM LAW FOR LOW FREQUENCIES}

\bigskip Let us consider the case of low frequencies where $w\ll $ $%
\overline{\Omega _{\parallel }^{2}}_{e\left( i\right) }\tau _{\ast }\>(%
\overline{\Omega _{\perp }^{2}}_{e\left( i\right) }\tau _{\ast })$ and take
into account the collision frequency $\nu =1/\tau _{c}\gg w$ ($1/\tau
_{c}^{\prime }$ for ions). To simplify our calculations, we also assume a
weak anisotropy of $\overline{\Omega _{\perp }^{2}}_{e\left( i\right)
}\approx $ $\overline{\Omega _{\parallel }^{2}}_{e\left( i\right) }$ and $%
\alpha _{\perp e\left( i\right) }\lambda _{\kappa \perp }\approx \alpha
_{\parallel e\left( i\right) }\lambda _{\kappa \parallel }=\alpha _{e\left(
i\right) }\lambda _{\kappa }$ .

The conductivity tensor (\ref{sigma}) in this limit will then appear as
\begin{gather}
4\pi \widehat{\sigma }_{kl}(\mathbf{k},w)=\left( \frac{\omega _{e}^{2}\tau
_{e}}{1+\Omega _{e}^{2}\tau _{e}^{2}}+\frac{\omega _{i}^{2}\tau _{i}}{%
1+\Omega _{i}^{2}\tau _{i}^{2}}\right) \delta _{kl}+\left( \frac{\omega
_{e}^{2}\Omega _{e}^{2}\tau _{e}^{3}}{1+\Omega _{e}^{2}\tau _{e}^{2}}+\frac{%
\omega _{i}^{2}\Omega _{i}^{2}\tau _{i}^{3}}{1+\Omega _{i}^{2}\tau _{i}^{2}}%
\right) l_{k}l_{l}  \notag \\
-\left( \frac{\omega _{e}^{2}\Omega _{e}\tau _{e}^{2}}{1+\Omega _{e}^{2}\tau
_{e}^{2}}-\frac{\omega _{i}^{2}\Omega _{i}\tau _{i}^{2}}{1+\Omega
_{i}^{2}\tau _{e}^{2}}\right) \varepsilon _{kml}l_{m}  \notag \\
+i\left( \alpha _{e}\frac{\omega _{e}^{2}\Omega _{e}\tau _{e}^{2}}{1+\Omega
_{e}^{2}\tau _{e}^{2}}\left( 1+\frac{\tau _{e}}{\tau _{c}}\right) -\alpha
_{i}\frac{\omega _{i}^{2}\Omega _{i}\tau _{i}^{2}}{1+\Omega _{i}^{2}\tau
_{i}^{2}}\left( 1+\frac{\tau _{i}}{\tau _{c}^{\prime }}\right) \right)
\lambda _{\kappa }\left( l_{m}k_{m}\delta _{kl}-l_{l}k_{k}\right)  \notag \\
+i\left( \alpha _{e}\omega _{e}^{2}\left( \tau _{c}+\tau _{e}\right) +\alpha
_{i}\omega _{i}^{2}\left( \tau _{c}^{\prime }+\tau _{i}\right) \right)
\lambda _{\kappa }l_{k}l_{m}\varepsilon _{mnl}k_{n}  \notag \\
+i\left( \alpha _{e}\frac{\omega _{e}^{2}\left( \tau _{c}+\tau _{e}\right) }{%
1+\Omega _{e}^{2}\tau _{e}^{2}}+\alpha _{i}\frac{\omega _{i}^{2}\left( \tau
_{c}^{\prime }+\tau _{i}\right) }{1+\Omega _{i}^{2}\tau _{i}^{2}}\right)
\lambda _{\kappa }\left( \varepsilon _{kml}k_{m}-l_{k}l_{m}\varepsilon
_{lmn}k_{n}\right)
\end{gather}
Here, we introduced the following characteristic time scales:
\begin{equation}
\tau _{e}=\frac{\tau _{c}}{1+\overline{\Omega _{\parallel e}^{2}}\tau _{\ast
}\tau _{c}},\tau _{i}=\frac{\tau _{c}^{\prime }}{1+\overline{\Omega
_{\parallel }^{2}}_{i}\tau _{\ast }^{\prime }\tau _{c}}  \label{tau}
\end{equation}
Having defined the conductivities
\begin{equation*}
\sigma _{0e\left( i\right) }=\frac{\omega _{e\left( i\right) }^{2}\tau
_{e\left( i\right) }}{4\pi },\sigma _{e\left( i\right) \perp }=\frac{\sigma
_{0e\left( i\right) }}{1+\Omega _{e\left( i\right) }^{2}\tau _{e\left(
i\right) }^{2}};\sigma _{e\left( i\right) \parallel }=\frac{\sigma
_{0e\left( i\right) }\Omega _{e\left( i\right) }^{2}\tau _{e\left( i\right)
}^{2}}{1+\Omega _{e\left( i\right) }^{2}\tau _{e\left( i\right) }^{2}},
\end{equation*}%
the current after applying the inverse Fourier transform can be written as
\begin{equation}
\begin{array}{c}
\mathbf{j}=\left( \sigma _{e\perp }+\sigma _{i\perp }\right) \left\langle
\mathbf{E}\right\rangle +\left( \sigma _{e\parallel }+\sigma _{i\parallel
}\right) \mathbf{l}\left( \mathbf{l}\left\langle \mathbf{E}\right\rangle
\right) -\left( \sigma _{e\perp }\Omega _{e}\tau _{e}-\sigma _{i\perp
}\Omega _{i}\tau _{i}\right) \left[ \mathbf{l}\left\langle \mathbf{E}%
\right\rangle \right] \\
+\left( \alpha _{e}\sigma _{e\perp }\Omega _{e}\tau _{e}\left( 1+\frac{\tau
_{e}}{\tau _{c}}\right) -\alpha _{i}\sigma _{ie\perp }\Omega _{i}\tau
_{i}\left( 1+\frac{\tau _{i}}{\tau _{c}^{\prime }}\right) \right) \lambda
_{\kappa }\left( \left( \mathbf{l\nabla }\right) \left\langle \mathbf{E}%
\right\rangle -\mathbf{\nabla }\left( \mathbf{l}\left\langle \mathbf{E}%
\right\rangle \right) \right) \\
+\left( \alpha _{e}\left( \sigma _{0e}-\sigma _{e\perp }\right) \left( 1+%
\frac{\tau _{c}}{\tau _{e}}\right) +\alpha _{i}\left( \sigma _{0i}-\sigma
_{i\perp }\right) \left( 1+\frac{\tau _{c}^{\prime }}{\tau _{i}}\right)
\right) \lambda _{\kappa }\mathbf{l}\left( \mathbf{l}\text{rot}\left\langle
\mathbf{E}\right\rangle \right) \\
+\left( \alpha _{e}\sigma _{e\perp }\left( 1+\frac{\tau _{c}}{\tau _{e}}%
\right) +\alpha _{i}\sigma _{i\perp }\left( 1+\frac{\tau _{c}^{\prime }}{%
\tau _{i}}\right) \right) \lambda _{\kappa }\text{rot}\left\langle \mathbf{E}%
\right\rangle%
\end{array}
\label{om}
\end{equation}
The influence of fluctuations and external magnetic field primarily causes
the conductivity to decrease, while the presence of helicity leads to an
additional dependence of the current on the vortex component of the electric
field. Let us consider the mean magnetic field in a medium with the Ohm law (%
\ref{om}). We will disregard the ion component (the EMHD approximation).
Neglecting the displacement current, we obtain the following expression for
the growth rate $\gamma $ of wave field perturbations of the form
\begin{equation*}
\begin{array}{c}
\left\langle \mathbf{E}\right\rangle =\exp \left( \gamma t\right) \left(
E_{x}\left( z\right) ,E_{y}\left( z\right) ,0\right) , \\
\left\langle \mathbf{B}\right\rangle =\exp \left( \gamma t\right) \left(
B_{x}\left( z\right) ,B_{y}\left( z\right) ,0\right)%
\end{array}%
\end{equation*}%
\begin{equation}
\gamma =-\frac{c^{2}\,k^{2}}{4\,\pi \,\sigma _{e\perp }}\frac{\,\left( \tau
_{c}+i\,\alpha _{e}k\,\lambda \,\,\zeta \,\Omega _{e}\tau _{e}\right) }{%
\,\left( \,\left( 1-i\,\tau _{e}\,\Omega _{e}\right) -\alpha _{e}k\,\lambda
\,\zeta \,\left( \tau _{c}/\tau _{e}-i\,\Omega _{e}\tau _{e}\,\right)
\right) \,\left( \,\left( 1+i\,\,\Omega _{e}\tau _{e}\right) +\alpha
_{e}k\,\,\lambda \,\zeta \,\left( \tau _{c}/\tau _{e}+i\,\,\Omega _{e}\tau
_{e}\right) \right) },
\end{equation}%
where $\zeta =1+\tau _{e}/\tau _{c}$. For the wave vectors
\begin{equation*}
\alpha _{e}\left\vert k\,\lambda \right\vert >\frac{\left( 1+\Omega
_{e}^{2}\tau _{e}^{2}\right) ^{1/2}}{\left( 1+\tau _{e}/\tau _{c}\right)
\left( \tau _{c}^{2}/\tau _{e}^{2}+\left( 2\tau _{c}/\tau _{e}-1\right)
\Omega _{e}^{2}\tau _{e}^{2}\right) ^{1/2}}
\end{equation*}%
the perturbations grow. For intense magnetic fluctuations, $\tau _{e}\approx
1/\left( \overline{\Omega _{\parallel }^{2}}\tau _{\ast }\right) \ll \tau
_{c}$ , and we obtain for the threshold wave number\bigskip
\begin{equation*}
\alpha _{e}\left\vert k\,\lambda \right\vert \gtrsim \frac{1}{\overline{%
\Omega _{\parallel }^{2}}\tau _{\ast }\tau _{c}}\left( 1+\frac{\Omega
_{e}^{2}}{2\left( \overline{\Omega _{\parallel }^{2}}\tau _{\ast }\right)
^{2}}\right) .
\end{equation*}%
In the collisionless limit $\tau _{c}\rightarrow 0$ , the threshold wave
number is
\begin{equation*}
\alpha _{e}\left\vert k\,\lambda \right\vert >\frac{1}{2}
\end{equation*}%
i.e., the threshold instability scale also increases with fluctuation
amplitude (parameter $\alpha _{e}$). On this threshold scale, the waves with
the following frequency propagate at $\tau _{e}\ll \tau _{c}$:
\begin{equation*}
w=\frac{c^{2}\,}{8\,\pi \,\sigma _{0e}}\frac{1}{\alpha _{e}^{2}\lambda ^{2}%
\overline{\Omega _{\parallel }^{2}}\tau _{\ast }\tau _{c}}\frac{1+\Omega
_{e}^{2}\tau _{c}^{2}}{\Omega _{e}\tau _{c}}\left( 1+\frac{\Omega _{e}^{2}}{%
2\left( \overline{\Omega _{\parallel }^{2}}\tau _{\ast }\tau _{c}\right) ^{2}%
}\right)
\end{equation*}%
Retaining the quadratic terms in the permittivity tensor gives rise to
dissipative terms of the form $-\sigma _{\ast }\Delta \left\langle \mathbf{E}%
\right\rangle +\sigma _{\ast }^{\prime }\nabla $div$\left\langle \mathbf{E}%
\right\rangle $ [10] in the Ohm law. Their influence restricts the
instability region, and the field perturbations are damped on small scales.

\section{FINITE CORRELATION TIMES}

\bigskip Consider the effects of finite correlation times for high
frequencies, $w\tau \gg 1$, with the anisotropy effects disregarded. These
also include the case of long correlation times. In this limit, the
effective Lorentz force is
\begin{gather*}
\frac{e}{mc}\int \left\langle \widehat{\mathbf{v}}(\mathbf{q},s\mathbf{%
)\times }\widehat{\mathbf{B}}(\mathbf{k-q},w-s)\right\rangle d\mathbf{q}ds=-%
\frac{4}{3}iw\tau \left( \frac{e}{mc}\right) ^{2}\widehat{\mathcal{E}}\tau
_{\ast }\left\langle \widehat{\mathbf{v}}(\mathbf{k},w)\right\rangle \\
-\frac{2\tau }{3}\frac{e}{m}\left( \frac{e}{mc}\right) ^{2}H_{0\perp }\tau
_{\ast }i\left[ \mathbf{k\times }\left\langle \widehat{\mathbf{E}}(\mathbf{k}%
,w)\right\rangle \right] \\
+\frac{2}{3}\left( \frac{e}{mc}\right) ^{2}E_{0}\tau _{\ast }\tau \left[
\mathbf{\Omega _{e}}\times \left\langle \widehat{\mathbf{v}}(\mathbf{k}%
,w)\right\rangle \right] -\frac{2}{3}i\tau w\left( \frac{e}{mc}\right)
^{2}H_{0}\tau _{\ast }\mathbf{\Omega _{e}}\delta ({\mathbf{k}})\delta (w),
\end{gather*}%
The gyrotropic fluctuational acceleration will be replaced with
oscillations. The frequencies will acquire a negative shift, and the
permittivity tensor will take the form
\begin{equation}
\widehat{\varepsilon }=\left(
\begin{array}{ccc}
\varepsilon _{\perp }-w\tau \chi _{0}k_{z} & ig+iw\tau \chi _{\perp }k_{z} &
w\tau \chi _{0}k_{x}-iw\tau \chi _{\perp }k_{y} \\
-ig-iw\tau \chi _{\perp }k_{z} & \varepsilon _{\perp }-w\tau \chi _{0}k_{z}
& w\tau \chi _{0}k_{y}+iw\tau \chi _{\perp }k_{x} \\
iw\tau \chi _{\parallel }k_{y} & -iw\tau \chi _{\parallel }k_{x} &
\varepsilon _{\parallel }%
\end{array}%
\right) .
\end{equation}%
Consider the waves propagating along the magnetic field, $\theta =0$. In
this case, the dispersion relation (\ref{dispeq}) has the solutions
\begin{equation}
\begin{array}{c}
n_{1,2}=\left( \varepsilon _{\perp }+g\right) \left( 1\pm \frac{2w\tau
\left( \chi _{0}-\chi _{\perp }\right) }{\left( 4c^{2}\left( \varepsilon
_{\perp }+g\right) ^{2}+\left( \chi _{0}-\chi _{\perp }\right) ^{2}w^{4}\tau
^{2}\right) ^{1/2}\mp \left( \chi _{0}-\chi _{\perp }\right) cw^{2}\tau }%
\right) \\
n_{3,4}=\left( \varepsilon _{\perp }-g\right) \left( 1\pm \frac{2w^{2}\tau
\left( \chi _{0}+\chi _{\perp }\right) }{\left( 4c^{2}\left( \varepsilon
_{\perp }-g\right) ^{2}+\left( \chi _{0}-\chi _{\perp }\right) ^{2}w^{4}\tau
^{2}\right) ^{1/2}\mp \left( \chi _{0}+\chi _{\perp }\right) cw^{2}\tau }%
\right)%
\end{array}%
\end{equation}
Assuming the helical additions to be small, we can write
\begin{eqnarray}
n_{1,2} &=&\left( \varepsilon _{\perp }+g\pm \frac{w^{2}\tau \left( \chi
_{0}-\chi _{\perp }\right) }{c}\right) =  \notag \\
&&1-\frac{\omega _{e}^{2}}{w\left( w-\Omega _{e}\right) }-\frac{\omega
_{i}^{2}}{w\left( w+\Omega _{i}\right) }\mp \left( \alpha _{\perp e}\frac{%
\omega _{e}^{2}\Omega _{e}\tau }{\left( w+\Omega _{e}\right) \Omega _{\kappa
\perp }}+\alpha _{\perp i}\frac{\omega _{i}^{2}\Omega _{i}\tau }{\left(
w-\Omega _{i}\right) \Omega _{\kappa \perp }}\right) \\
n_{3,4} &=&\left( \varepsilon _{\perp }-g\pm \frac{w^{2}\tau \left( \chi
_{0}+\chi _{\perp }\right) }{c}\right) =  \notag \\
&&1-\frac{\omega _{e}^{2}}{w\left( w+\Omega _{e}\right) }-\frac{\omega
_{i}^{2}}{w\left( w-\Omega _{i}\right) }\pm \left( \alpha _{\perp e}\frac{%
\omega _{e}^{2}\Omega _{e}\tau }{\left( w-\Omega _{e}\right) \Omega _{\kappa
\perp }}+\alpha _{\perp i}\frac{\omega _{i}^{2}\Omega _{i}\tau }{\left(
w+\Omega _{i}\right) \Omega _{\kappa \perp }}\right)
\end{eqnarray}
It is easy to see that an additional rotation of the polarization plane
appears here.

For the waves propagating perpendicular to the magnetic field, $\theta =%
\frac{\pi }{2}$, we obtain the following solutions:%
\begin{equation}
\begin{array}{c}
n_{1}^{2}=\frac{\varepsilon _{\perp }\left( \varepsilon _{\perp
}+\varepsilon _{\parallel }\right) -g^{2}+\kappa ^{\prime }+\left( \left(
\varepsilon _{\perp }\left( \varepsilon _{\perp }-\varepsilon _{\parallel
}\right) -g^{2}\right) ^{2}-2\left( \varepsilon _{\perp }\left( \varepsilon
_{\perp }+\varepsilon _{\parallel }\right) -g^{2}\right) \kappa ^{\prime
}+\kappa ^{\prime 2}\right) ^{1/2}}{2\varepsilon _{\perp }} \\
n_{2}^{2}=\frac{\varepsilon _{\perp }\left( \varepsilon _{\perp
}+\varepsilon _{\parallel }\right) -g^{2}+\kappa ^{\prime }-\left( \left(
\varepsilon _{\perp }\left( \varepsilon _{\perp }-\varepsilon _{\parallel
}\right) -g^{2}\right) ^{2}-2\left( \varepsilon _{\perp }\left( \varepsilon
_{\perp }+\varepsilon _{\parallel }\right) -g^{2}\right) \kappa ^{\prime
}+\kappa ^{\prime 2}\right) ^{1/2}}{2\varepsilon _{\perp }} \\
\kappa ^{\prime }=w^{4}\tau ^{2}\left( g\chi _{0}+\varepsilon _{\perp }\chi
_{\perp }\right) \chi _{\parallel }/c^{2}%
\end{array}%
\end{equation}

As in the approximation of a $\delta $-correlated random process (\ref{Bperp}%
) considered above, the propagation conditions change, and elliptical
polarization attributable to helicity appears in both the ordinary and
extraordinary waves. Note that in the case of infinite correlation times or
high frequencies (frozen fluctuations), the properties of a plasma medium
with magnetic helicity are similar to those of chiral and bianisotropic
media [14, 15]. In real systems, $w\tau $ is finite, and the effects of both
instability considered in Section 4 and the appearance of additional wave
modes must simultaneously manifest themselves.

\section{CONCLUSIONS}

\bigskip The influence of magnetic fluctuations on the motion of the
particles of a cold magnetoactive plasma primarily reduces to the appearance
of an effective fluctuational collision frequency determined by the
statistical parameters and to the decrease in conductivity. Reflectional
symmetry breaking---nonzero mean magnetic helicity of the
fluctuations---leads to a change in the dispersion of the propagating waves
and the appearance of additional modes. The waves can be unstable,
reflecting both the nonequilibrium nature of the turbulent magnetic helicity
and the peculiarities of the particle motion in random helical magnetic
fields. The instability growth rate is proportional to the helicity of the
fluctuational magnetic field and the amplitude of the large scale uniform
magnetic field. An allowance for the finite correlation times and for the
additional fluctuational quadratic dispersion effects restricts the action
of this instability. In contrast to the turbulent dynamo effects considered
in the MHD and EMHD approximations [2], here there is a natural restriction
of the instability region on large scales determined by the relationship
between the fluctuational helicity and energy and the large-scale magnetic
field. The plasma acquires properties similar to those observed in chiral
and bianisotropic media [14, 15]. Consequently, it can have properties
characteristic of these media, such as anomalous absorption [27, 28] and
additional wave conversion effects [29, 30]. In contrast to the artificial
external origin of the chirality in chiral media, this property is natural
in a turbulent magnetoactive plasma with helicity. The deviations in the
rotation of the polarization plane attributable to fluctuational helicity
can serve as a tool for diagnosing it. The results were obtained in the
approximation of isolated particles whose advantages and disadvantages are
well known. It is easy to see that the above effects are preserved when the
thermal and collisional effects are taken into account and can be obtained
in terms of the kinetic approach.

\bigskip I am grateful to S.N. Artekha and N.S. Erokhin for helpful
discussions. This work was supported by the Russian Science Support
Foundation.


\begin{thebibliography}{99}
\bibitem{Taylor74} J.B.Taylor, Phys.Rev.Lett. \textbf{33}, 1139 (1974).

\bibitem{Vainshtein80} S. I. Vainshte n, Ya. B. Zel'dovich, and A.A.Ruzma
kin, \textsl{The Turbulent Dynamo in Astrophysics} (Nauka, Moscow, 1980) [in
Russian].

\bibitem{Hasselman} K.Hasselmann, G. Wibberenz, Zs. Geophys. \textbf{34},
353 (1968).

\bibitem{Goldstein} M. L.Goldstein, W. H. Matthaeus, Proc. 17th Internat.
Cosmic Ray Conf., \textbf{3}, 294 (1981).

\bibitem{Kichatinov83} L. L. Kichatinov, Pis'ma Zh. Eksp. Teor. Fiz. \textbf{%
37}, 43 (1983) [JETP Lett. \textbf{37}, 51 (1983)].

\bibitem{Fedorov} Yu.I.Fedorov, V.E.Katz, L.L.Kichatinov \& M.Stehlic,
Astron. and Astrophys \textbf{260}, 499 (1992).

\bibitem{Mett} R.R.Mett, J.A.Tataronis, Phys.Rev.Lett. \textbf{63}, 1380
(1989).

\bibitem{Taylor} J.B.Taylor, Phys.Rev.Lett. \textbf{63}, 1384 (1989).

\bibitem{Tur94} A.V.Chechkin, V.V.Yanovsky and A.V.Tur, Phys.Plasmas \textbf{%
1}, 2566 (1994).

\bibitem{Chiral99} L. L. Kichatinov, Pis'ma Zh. Eksp. Teor. Fiz. 37, 43
(1983) [JETP Lett. 37, 51 (1983)].

\bibitem{Rukhadze} V. L. Ginzburg and A. A. Rukhadze, \textsl{Waves in
Magnetized Plasmas }(Nauka, Moscow, 1975) [in Russian].

\bibitem{Vainshtein98} S.I.Vaihnstein, Phys.Rev.Lett. \textbf{80}, 4879
(1998).

\bibitem{Aleksandrov} A. F. Aleksandrov, and A. A. Rukhadze, \textsl{%
Lectures on Electrodynamics of Plasma-like Media} (Mosk.Gos. Univ., Moscow,
1999) [in Russian].

\bibitem{Jaggard} D.L.Jaggard, A.R.Mickelson , C.H.Papas, Applied Physics
\textbf{18}, 211, (1979).

\bibitem{UFN} B. Z. Katsenelenbaum, E. N. Korshunova, A. N. Sizov, and A. D.
Shatrov, Usp. Fiz. Nauk \textbf{167}, 1201 (1997) [Phys. Usp. \textbf{40},
1149(1997)].

\bibitem{Novikov64} E. A. Novikov, Zh. Eksp. Teor. Fiz. \textbf{47}, 1919
(1964) [Sov. Phys. JETP \textbf{20}, 1290 (1964).

\bibitem{Klyatskin} V. I. Klyatskin and V. I. Tatarski , Izv. Vyssh. Uchebn.
Zaved., Radiofiz. \textbf{15}, 1433 (1972).

\bibitem{Vainshtein70} S. I. Vainshtein, Zh. Eksp. Teor. Fiz. \textbf{58},
153 (1970) [Sov. Phys. JETP\textbf{\ 31}, 87 (1970)].

\bibitem{Matthaeus81} W.H.Matthaeus, C.Smith, Phys.Rev. A \textbf{24}, 2135
(1981).

\bibitem{Radler97} S.Oughton, K.--H.R\"{a}dler, W.H.Matthaeus, Phys.Rev. E
\textbf{56}, 2875 (1997).

\bibitem{Krause80} F. Krause and K.-H. Radler, \textsl{Mean-Field
Magnetohydrodynamics and Dynamo Theory} (Akademie, Berlin, 1980; Mir,
Moscow, 1984).

\bibitem{Toptygin} I. N. Toptygin, \textsl{Cosmic Rays in Interplanetary
Magnetic Fields} (Nauka, Moscow, 1983; Reidel, Dordrecht, 1985).

\bibitem{Yanovsky} A.V.Chechkin, D.P.Sorokin, V.V.Yanovsky,
http://xxx.lanl.gov/ps/hep-th/9306159

\bibitem{Smith} C. W. Smith, J. W. Bieber, in\textit{\ Solar Wind Eight,}
AIP Conference Proceedings \textbf{382}, 498, AIP, New York (1996).

\bibitem{Soloviev} A. I. Morozov and L. S. Solov'ev, Zh. Tekh. Fiz. \textbf{%
30}, 271 (1960) [Sov. Phys. Tech. Phys. \textbf{5}, 250 (1960)].

\bibitem{Khantadze} A. G.Khantadze, G. D. Aburdzhania, G. V. Dzhandieri, and
Kh. Z. Chargaziya, Fiz.Plazmy (Moscow) \textbf{30}, 88 (2004) [Plasma Phys.
Rep. \textbf{30}, 83 (2004)].

\bibitem{Jaggard89} D.L.Jaggard, N.Engheta, Electronics Lett. \textbf{25},
1060 (1989).

\bibitem{Osipov99} V. A. Neganov and O. V. Osipov, Izv. Vyssh. Uchebn.
Zaved., Radiofiz. \textbf{42}, 870 (1999).

\bibitem{Erokhin} N.S.Erokhin, S.S.Moiseev, \textsl{On the mode conversion
in weakly inhomogeneous chiral plasma}, Preprint No. 1948, IKI AN RAN (Inst.
for Space Research, Russian Academy of Sciences, Moscow, 1996).

\bibitem{Silva} H.T.Torres, P.H.Sakanaka, N.Reggiani, J.Phys.Soc. of Japan
\textbf{67}, 850 (1998).
\end{thebibliography}
\end{document}